# Connecting bond switching to fracture toughness of calcium aluminosilicate glasses

Sidsel Mulvad Johansen[1], Tao Du[1,2], Johan F. S. Christensen[1], Anders K. R. Christensen[1], Xuan Ge[1,3], Theany To[4,5], Lars R. Jensen[6], Morten M. Smedskjaer[1,*]

[1]Department of Chemistry and Bioscience, Aalborg University, 9220 Aalborg East, Denmark

[2]Department of Applied Physics, The Hong Kong Polytechnic University, Kowloon, Hong Kong, 999077, China

[3]Shanghai Key Laboratory of Materials Laser Processing and Modification, School of Materials Science and Engineering, Shanghai Jiao Tong University, 200240 Shanghai, China

[4]Univ Rennes, CNRS, IPR (Institut de Physique de Rennes) - UMR 6251, Rennes, France

[5]Nantes Université, Ecole Centrale Nantes, CNRS, GeM, UMR 6183, Nantes F-44000, France

[6]Department of Materials and Production, Aalborg University, 9220 Aalborg East, Denmark

* Corresponding author. E-mail: mos@bio.aau.dk

## Abstract

Fracture toughness is a critical mechanical property of glasses, but a detailed understanding of its link to composition and structure is still missing. Here, focusing on the industrially important family of calcium aluminosilicate glasses, we measure the fracture toughness of two glass series using the single-edge precracked beam method, one based on tectosilicate compositions with varying silica contents and the other covering both percalcic and peraluminous compositions with varying Al/Ca ratio. To elucidate the structural origins of the variation in fracture toughness, we perform X-ray total scattering measurements and molecular dynamics simulations. Our findings show that local coordination changes of especially Al atoms, so-called bond switching, feature an overall positive correlation with fracture toughness. We also compare this variation with that in other mechanical properties, including elastic moduli, hardness, and crack initiation resistance. We find that various structural aspects need to be considered to describe and understand the mechanical properties of calcium aluminosilicate glasses.

# 1. Introduction

Oxide glasses are used in various applications, including windows, screens on electronic devices, tableware, containers and optical devices or fibers. This is due to their flexibility in terms of chemical composition and easy formability, but also their favorable properties such as high hardness and transparency[1,2]. Oxide glasses are by far the most common glass type as the raw materials are typically readily available, with a drawback being the high melting point[2]. Another major drawback is their brittleness. That is, their disordered structure combined with strong and highly directional covalent bonds prevent the glasses from undergoing significant plastic deformation at the macroscale. This causes them to fracture at low tensile loads as a result of stress concentrations at surface flaws and their inability to undergo plastic flow, which is in contrast to, e.g., metals with non-directional bonds[1]. If the fracture resistance of oxide glasses could be improved, their lifetime and safety would be enhanced, expanding the range of potential applications and allowing the production of thinner glass products. Despite the importance of understanding glass brittleness, we still know relatively little about the relation between the mechanical properties and the underlying glass structure, largely due to the non-crystalline nature of glass. To this end, it is important to investigate both the resistance to formation of cracks (e.g., surface flaws) and the resistance to crack propagation.

The resistance to crack formation in glasses is usually quantified through Vickers indentation tests[3] at different loads and by counting the number of corners with cracks. The crack resistance ($CR$), as proposed by Wada et al.[4], is then defined as the load with 50% crack initiation probability. Upon indentation, oxide glasses will deform elastically, as well as plastically through a combination of densification and shear flow. Each mechanism occurs to a varying degree for different glasses, resulting in different $CR$ values and crack morphologies.[1,5] Cracking occurs when sufficiently high tensile stresses are induced locally in the glass, e.g., when an indentation load is applied and causes ring cracking. Cracking can also occur during removal of the indentation load where tensile stresses are induced at the elastic-plastic boundary, as the plastically deformed material will not recover to its original shape[1,5,6]. "Normal" glasses (e.g., soda-lime silica glass) primarily deform through shear flow, whereas "anomalous" glasses (e.g., silica glass) primarily deform through densification[6,7]. Intermediate glasses (e.g., some aluminoborosilicates) deform with more densification than normal glasses and more shear flow than anomalous glasses, and these glasses can therefore feature higher crack resistance.[7,8]

The resistance to crack growth is typically quantified as the fracture toughness ($K_{Ic}$). The methods to accurately determine $K_{Ic}$ are time-consuming and require careful sample preparation. The indentation fracture toughness method is attractive, as it is the easiest and least time-consuming method, but it has been reported to be inaccurate.[9–11] Accurate self-consistent methods include single-edge precracked beam (SEPB) and chevron-notched beam (CNB) methods.[9,10] For these, a controlled precrack is introduced into the glass before measuring the resistance to crack growth, avoiding, e.g., the unwanted influence of densification when measuring indentation fracture toughness.[9,11–13] The composition and structure dependence of fracture toughness in glasses has been investigated using approaches based on, e.g., bond switching phenomena,[14,15] topological constraint theory (TCT)[16], and bond energies[10,17]. First, the bond switching approach considers the proposed energy dissipation during crack growth by the formation, breaking or rearrangement (swapping) of bonds. Some elements, such as Al and B, are more prone to undergo bond switching than others.[14,18] Second, TCT describes the dissipation of energy based on the transition between stressed-rigid and flexible regions, which is controlled by the chemical composition. The region between these two types, called the isostatic region, is the most resistant to crack propagation as it is characterized by being rigid but stress-free.[16] Third, Rouxel's bond-energy-based model predicts the fracture toughness from Young's modulus, the Poisson's ratio, and the fracture surface energy. The latter is calculated from bond energies and assumed type and number of broken bonds for crack growth along the easiest fracture pathway.[9,10,17] The bond switching theory and the TCT have shown to describe some experimental data well, but they require computational modelling[14,16,19]. Rouxel's model provides a relatively simple estimate of the fracture toughness without requiring computational modeling, but it may have limitations for glasses with cations that are prone to bond switching[14,17]. While these approaches aim to explain fracture toughness at the atomic scale, there is still a gap of knowledge of understanding the relationship between the structural properties and the fracture toughness of oxide glasses, especially due to the lack of reliable experimental data.

In this work, we focus on calcium aluminosilicate (CAS) glasses, which are known for their chemical durability, thermal stability, and hardness[20,21]. Despite this, the connection between the composition, structure, and mechanical properties of CAS glasses is yet to be described in detail. Nuclear magnetic resonance spectroscopy studies have described the coordination of Si to be four-fold coordinated ($Si^{IV}$) and the Al atoms to be a mixture of four-fold ($Al^{IV}$), five-fold ($Al^V$), and six-fold ($Al^{VI}$) depending on the composition[22–24]. When Al is present in four-fold coordination ($Al^{IV}$), these tetrahedra are charge-compensated by Ca (or oxygen triclusters, i.e., three-fold coordinated oxygens). The four-fold coordinated state is the preferred state of Al when Ca is available, and in the peraluminous region, where there is an excess of Al compared to Ca, the excess Al is present in the higher coordinated states ($Al^V$ and $Al^{VI}$). When calcium is in excess compared to aluminum (percalcic region), non-bridging oxygens (NBOs) form, whereas tectosilicate glasses ($Al_2O_3/CaO$ = 1) are in principle fully polymerized and only contain bridging oxygens (BOs). However, it is well established that there are small amounts of $Al^V$ and $Al^{VI}$ present as well as NBOs even when there is enough calcium ions present to charge compensate $Al^{IV}$.[22–24]

Regarding the mechanical properties of CAS glasses, hardness has been shown to increase when silica and calcia is exchanged by alumina[25–27]. Lamberson et al.[25] found that the hardness increases with increasing amounts of $Al^V$ and $Al^{VI}$ in tectosilicate glasses, and hardness also shows a steep increase in the peraluminous region[27], thus the $Al^V$ and $Al^{VI}$ species appear to be correlated with a higher hardness. The latter study also showed that *CR* increases with increasing alumina content but decreases with the amount of NBOs. This is supported by the work of Shan et al.[21], who found that the presence of NBOs reduces *CR*, while oxygen triclusters significantly increase *CR*, as they make the network more prone to energy dissipation through bond switching. Pönitzsch et al.[27] have shown that the indentation fracture toughness decreases when silica is substituted with alumina, while it increases with larger Al/Ca ratio. A similar study by Gross et al.[11] found that the fracture toughness (measured using the chevron notch short bar method) of tectosilicate glasses in the range from 50 to 80 mol% $SiO_2$ decreases with increasing amount of silica. Molecular dynamics (MD) simulations have found that higher alumina content increases the fracture toughness, fracture strain, and tensile strength of CAS glasses.[19] The simulated structures show that the fraction of oxygen triclusters increases significantly with the alumina content, in alignment with experimental findings.[23]

The above results suggest that hardness, crack resistance, and fracture toughness should all increase with the alumina content. However, while it is already well-established that hardness and crack resistance generally increase with the alumina content[21], the fracture toughness of low-silica tectosilicate glasses as well as percalcic and peraluminous glasses have not yet been investigated experimentally. Therefore, we here investigate two series of CAS glasses, including one with tectosilicate compositions of varying silica content and one with varying Al/Ca ratio covering both percalcic and peraluminous compositions. We compare the experimental fracture toughness data, as determined by SEPB, with other mechanical properties as well as structural features determined from both experiments and MD simulations. We find that bond switching activities of especially Al atoms show the strongest positive correlation with fracture toughness.

## 2. Methods

*2.1 Sample preparation*

Two series of five glasses were synthesized, with one glass being present in both series, i.e., a total of nine glasses. Series I is based on tectosilicate compositions with the silica content varying from 20 to 60 mol%: $20CaO-20Al_2O_3-60SiO_2$, $25CaO-25Al_2O_3-50SiO_2$, $30CaO-30Al_2O_3-40SiO_2$, $35CaO-35Al_2O_3-30SiO_2$ and $40CaO-40Al_2O_3-20SiO_2$. Series II was designed with constant silica content of 40 mol%, while the calcia is exchanged by alumina to obtain $Al_2O_3/CaO$ ratios from 0.3 to 1.3: $25CaO-35Al_2O_3-40SiO_2$, $30CaO-30Al_2O_3-40SiO_2$, $35CaO-25Al_2O_3-40SiO_2$, $40CaO-20Al_2O_3-40SiO_2$, and $45CaO-15Al_2O_3-40SiO_2$. Series I was designed to understand the effect of silica, i.e., based on the $K_{Ic}$ results of Gross et al.[11] but focusing on low-silica glasses. Series II was designed to contain glasses with an expected varying propensity for different bond switching mechanisms, since the Al cations are most prone to undergo bond swapping, whereas Ca cations are more prone to undergo increase and decrease in coordination numbers as found from the MD simulations. The sample names reflect the compositions as C$x$A$y$S$z$, where $x$ is the amount of calcia, $y$ the amount of alumina and $z$ the amount silica (in mol%).

The glasses were prepared by mixing $CaCO_3$ (Chemsolute, >99.5%), $Al_2O_3$ (Sigma Aldrich, >99.5%) and $SiO_2$ (Sigma Aldrich, >99.95%) powders. These were poured in Pt-Rh crucibles and melted in an electrical furnace (Entech) at around 1600-1700 °C and homogenized for three hours before quenching. To improve homogeneity, the quenched glasses were crushed, melted again, and homogenized for another three hours. After performing differential scanning calorimetry (DSC) (STA 449 F1, Netzsch) measurements to determine the glass transition temperature ($T_g$), the glasses were annealed at their respective $T_g$ (**Table 1**) for 30 min. The non-crystallinity of the samples was verified by powder X-ray diffraction (XRD) analysis (PANalytical, Empyrean XRD, equipped with a Cu K$\alpha_1$ X-ray source), while the densities were determined using Archimedes principle of buoyancy in ethanol (≥ 99.99%) at room temperature. The chemical compositions of the glasses were analyzed by inductively coupled plasma optical emission spectroscopy.

Based on the measured compositions and density, the oxygen packing density, $V_O$, was calculated according to the method of Gross et al.[11] through **Eqs. 1-3**:

$$V_O = n_O \cdot \overline{v_O} \qquad (1)$$

$$n_O = \frac{\sum_i X_i \cdot n_i \cdot N_A}{V_m} \qquad (2)$$

$$\overline{v_O} = \frac{\sum_j k_j \cdot l_j \cdot CN_j \cdot \frac{4}{3} \cdot \pi \cdot r_{O_j}^3}{\sum_j k_j \cdot l_j \cdot CN_j} \qquad (3)$$

Here, $n_O$ is the number of oxygen atoms per unit volume, $i$ represents the three glass forming oxides ($i$ = CaO, $Al_2O_3$, $SiO_2$), $X_i$ is the mole fraction of oxide $i$, $n_i$ is the number of oxygen atoms in oxide $i$, $N_A$ is Avogadro's number, and $V_m$ is the molar volume of the glass. The second term, $\overline{v_O}$, is the mean volume that one oxygen atoms occupies, where $j$ represents the network forming oxides ($j$ = $Al_2O_3$, $SiO_2$), $k_j$ is the mole fraction of each network forming oxide at their given coordination state, $l_j$ is the number of cations in each network forming oxide, $CN_j$ is the corresponding coordination number, and $r_{O_j}$ is the corresponding oxide radius for each network forming species, which was set to 1.306, 1.437 and 1.334 Å for $Si^{IV}$, $Al^{IV}$, and $Al^V$, respectively[11].

*2.2 Structural characterization*

To characterize the structures of the glasses and compare these to the MD simulations described below, we performed high energy X-ray diffraction experiments. These were conducted at beamline 3W1 of Beijing Synchrotron Radiation Facility (BSRF), using an incident X-ray beam of wavelength 0.2062 Å (60.128 keV). A large-area detector (Mercu 1717HS, 3072 × 3048 pixels of 139 μm × 139 μm CsI) was placed 214 mm downstream of the sample. The setup was calibrated using the diffraction pattern from polycrystalline $CeO_2$ powder. The investigated glass samples were ground to discs with thickness of 0.7 mm and a diameter of approximately 10 mm and glued to an aluminum alloy frame with very thin Kompton tapes. The measurement procedure was controlled by detector software, and four repetitions of 5 s exposure were used for all samples. Background patterns were collected with the same set up and exposure time.

The obtained 2D detector data were azimuthally integrated using the pyFAI software[28] and the calibration with the $CeO_2$ powder. Based on this, the X-ray structure factor $S(Q)$ and pair distribution function (PDF) were calculated as follows. The scattering data was processed by GudrunX[29] and normalized to $<F>^2$, using the Breit-Dirac factor of 2.5, and with the Krogh-Moe and Norman normalization enabled, using the obtained elemental composition and estimated densities, resulting in a $Q_{max}$ of 22.2 Å$^{-1}$. To further correct the background, the contribution for unphysical peaks located at low $r$ (<1 Å) was suppressed. Moreover, the Top-hat function width was set to 1.5 Å$^{-1}$ as an additional filter for the Fourier Transform (FT) to real space. To minimize termination ripples, from the finite $Q$ ($Q_{max}$), a Lorch-like modification function was applied in the FT to real space (width of broadening 0.1 Å and broadening power 0.1).

*2.3 Mechanical characterization*

To test the mechanical properties, all samples were cut in different dimensions as required for each testing method. After cutting, the samples were first ground in water using SiC grinding papers and then polished to an optical finish using water-free polycrystalline diamond suspensions.

For elastic moduli measurements, samples with two parallel sides of approximately 2×2 cm$^2$ were prepared with an average thickness of 3.4 mm and ground on SiC grinding papers of grit size up to 2400. Using ultrasonic echography (38 DL Plus, Olympus), the velocity of sound waves through the glass were measured. By measuring the velocities of both longitudinal and transversal sound waves, the Young's modulus ($E$), the Poisson's ratio ($v$), the shear modulus ($G$) and the bulk modulus ($B$) were calculated based on the relations for isotropic materials[30].

Vickers indentation (Duramin-40 AC3, Struers) was used to determine the Vickers hardness ($H_V$) and crack resistance ($CR$) under laboratory conditions of 21.5±1.0 °C and 60±10% relative humidity. The measurements were performed on the samples also used for the echography measurements, but the surfaces used for indentation were polished to an optical finish prior to indentation. 20 to 30 indentations were made for each selected load, which was varied between 50 gf (0.49 N) and 2 kgf (19.61 N), with a dwell time of 10 s. To determine $CR$, the number of corners with redial cracks were counted after waiting for ~24 h. The crack initiation probability ($CP$) was then calculated as the average number of corners with cracks relative to the total of four corners for the Vickers indenter. By using a logarithmic sigmoid fit ($CP = 100/(1 + \left(\frac{P}{CR}\right)^{-K}$), $CR$ was determined as the load at which the crack probability was 50%.[4,8] The Vickers hardness was calculated as,

$$H_V = 1.854 \cdot \frac{P}{d^2} \qquad (4)$$

where *P* is the indentation load and *d* is the average diagonal length of the indentation imprint using an optical microscope. We determined the hardness at the load of 200 gf (1.96 N) to minimize indentation size effect[31,32], including only indents without cracks.

To evaluate the indentation deformation mechanism, i.e., the amount for densification during indentation, the side length recovery ratio ($L_{SR}$) was measured[8,33]. Around 20 indents were made for each composition at a load of 200 gf (1.96 N) to avoid cracks and using a dwell time of 10 s. After acquiring optical micrographs of the indents, the samples were annealed at their respective $0.9T_g$ (scaled in Kelvin) for two hours, before the indents were imaged again. The difference of the side lengths before and after annealing was then measured. **Figure 1** shows an example of optical micrographs used for measurements of indent side length recovery on the C30A29S41 glass sample before (a) and after (b) annealing for two hours. $L_{SR}$ for each composition was then calculated as,

$$L_{SR} = \frac{L_{Si} - L_{Sf}}{L_{Si}} \qquad (5)$$

where $L_{Si}$ is the side length before annealing and $L_{Sf}$ is the side length after annealing. Indents with developed cracks were discarded from this analysis, resulting in 13 to 21 indents being measured for each composition.

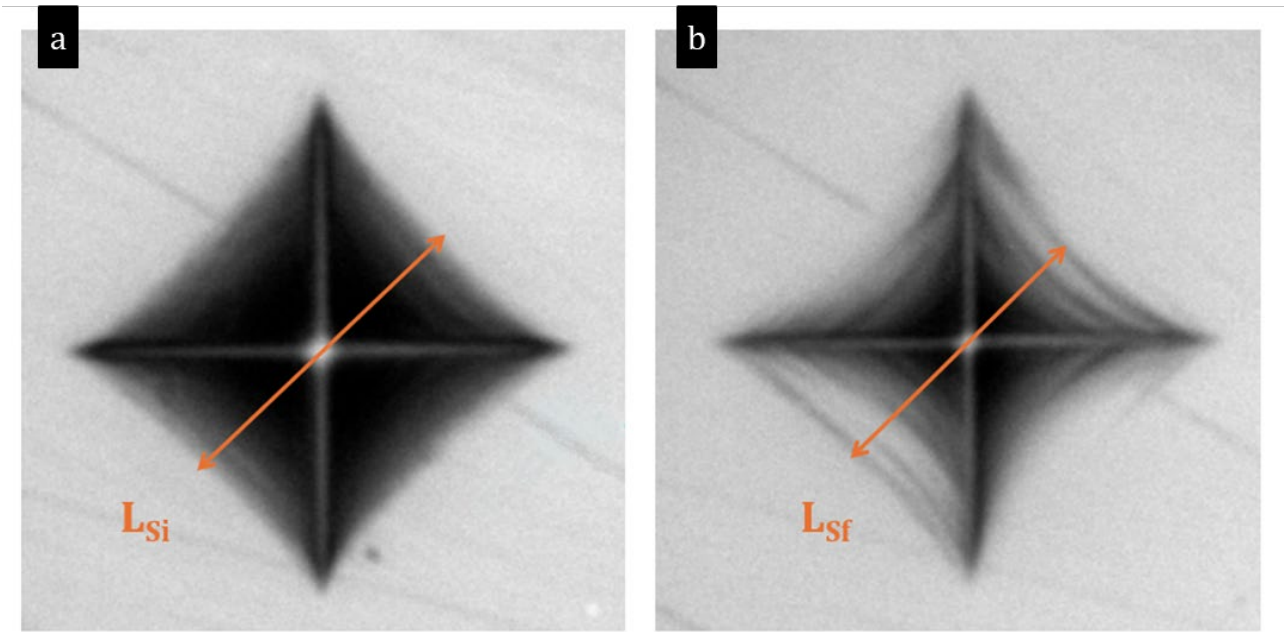

**Figure 1**. Example of optical micrographs used for measurements of indent side length recovery on the C30A29S41 glass sample (a) before and (b) after annealing at $0.9T_g$ for two hours. The indent was made using an indentation load of 200 gf and $L_{Si}$ was measured to be 14.9 µm whereas $L_{Sf}$ was measured to be 14.7 µm.

For the fracture toughness testing, the single-edge precracked beam (SEPB) method was performed according to the ASTM standard C1421-18, following the modified set-up for glasses [9]. To this end, glass beams were cut and polished into dimensions of 21×4×3 mm³ ($L \times W \times B$), where the planar variance across the surfaces of the beams was <30 µm. To guide the precrack formation, a line of indents were made in the middle of the beam (perpendicular to the direction of the length $L$) on the side of 3 mm width ($B$) (**Figure 2a**), with an applied indentation load around the respective $CR$ value and a dwell time of 15 sec. The indents were positioned carefully to connect the corner cracks from the indents while seeking to keep the formation of lateral cracks to a minimum. By utilizing a universal testing machine (Z100, Zwick), bridge compression was then performed between 12-24 hours after the indentation using a tungsten carbide setup with a 7 mm groove to make a controlled precrack (**Figure 2b**). The precrack was guided from the indentation line on side $B$ to approximately half the height of $W$, with a loading rate of 0.05 mm/min to ensure controlled precrack growth. Immediately thereafter, three-point bending was performed to fracture the precracked beam using a 1 kN load cell (instead of the 50 kN in the precracking step) for the reason of accuracy on the applied load (around 10 N to fracture the beam). The beam was placed with the precrack facing downwards on a three point bending fixture, with a distance between the rods of 16 mm (the so-called span width $S$), and the sample was then fractured using a loading rate of 0.9 mm/min (**Figure 2c**). The precrack length was then measured in the post-mortem observation of the fracture surface of the broken beam using an optical microscope (Nanovea). The precrack length relative to $W$ was measured at three locations of 25%, 50% and 75% relative to $B$ (**Figure 2d**). Only samples with precrack height deviations ≤15% between the three measured lengths and an average length ($a$) within 35-60% were included in the calculations. The fracture toughness was then calculated from the load at which the beam fractured (the peak load, $P_{\text{max}}$), $S, W, B$ and $a$ as,

$$K_{\text{Ic}} = \frac{P_{\text{max}}}{B\sqrt{W}} Y^* \quad (6)$$

$$Y^* = \frac{3}{2}\frac{S}{W}\frac{\alpha^{\frac{1}{2}}}{(1-\alpha)^{\frac{3}{2}}} f(\alpha) \quad (7)$$

Here, $\alpha$ is the $a/W$ ratio, and $f(\alpha) = [1.99 - (\alpha - \alpha^2)(2.15 - 3.93\alpha + 2.7\alpha^2)]/(1 + 2\alpha)$. The reported $K_{\text{Ic}}$ values were calculated as the average of three to eight successful individual measurements.

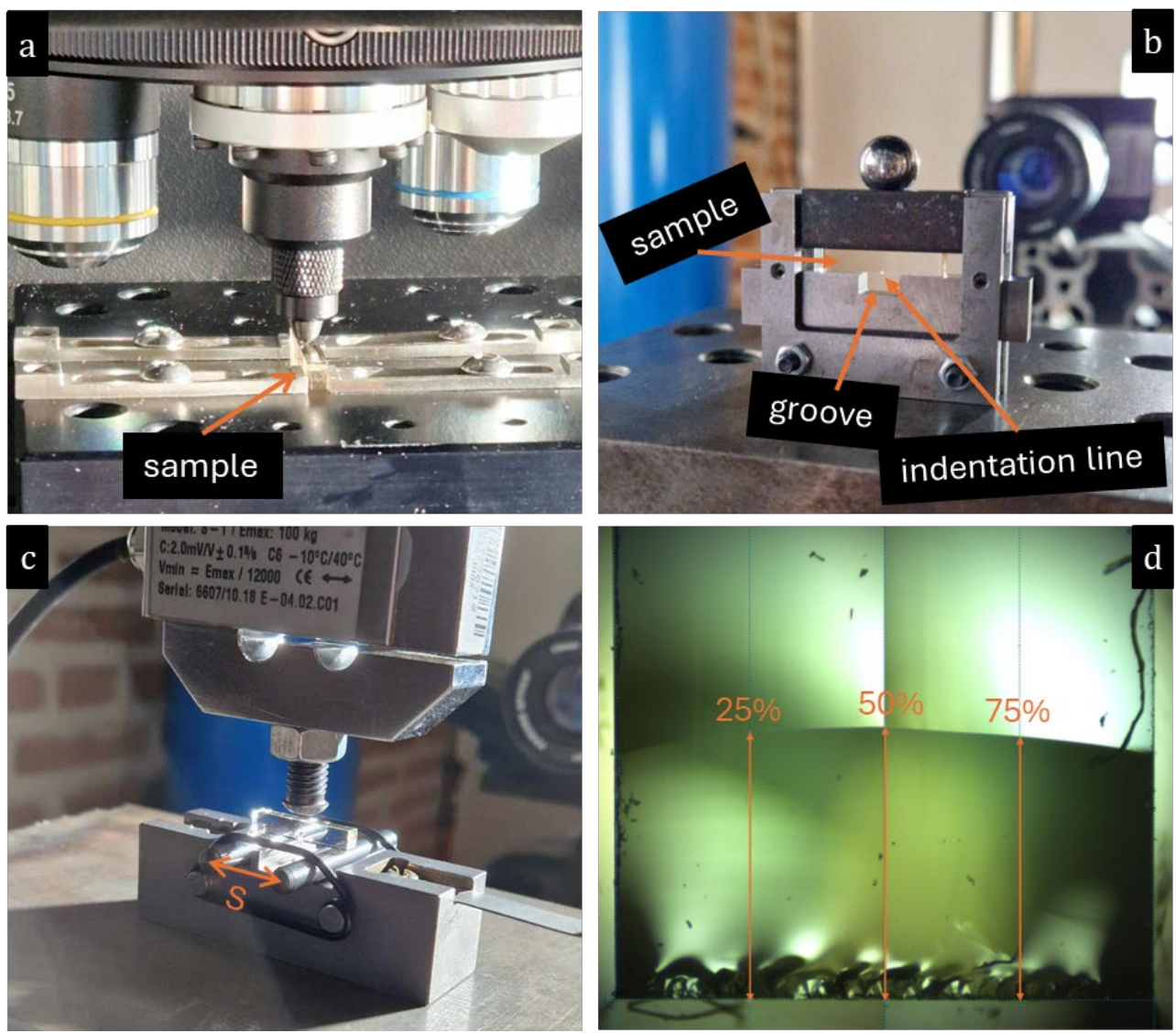

**Figure 2.** Single-edge precracked beam method to determine fracture toughness. (a) Indentation setup using a Vickers indenter to create a line of indents across the sample. (b) Bridge compression setup with a groove of 7 mm used to grow a precrack from the indentation line. (c) Three point bending setup with a span width (*S*) of 16 mm. (d) Example of a fractured beam and measurement of the precrack length at three different locations (25%, 50% and 75%) of the width, based on which the average precrack length is calculated.

*2.4 MD simulations*

The melt-quenching procedure to generate the same CAS glasses as in the experiments follows refs.[19,34]. Initial configurations of the CAS systems, containing approximately 3000 atoms, were generated by randomly placing the designated numbers of calcium, aluminum, silicon, and oxygen atoms into a cubic box using PACKMOL[35], while avoiding unrealistic overlaps by applying a cutoff distance of 2.0 Å. The interatomic interactions were described using a classical potential parameterized by Bouhadja et al.[36], which has been shown to reasonably reproduce the structural and mechanical properties of CAS glasses[20]. This potential adopts the Born–Mayer–Huggins interatomic interaction form with a short-range cutoff of 8.0 Å, while Coulomb interactions were calculated using the Wolf summation method with a damping parameter of 0.25 and a cutoff of 10 Å[37]. Atomic motion was integrated using the velocity–Verlet algorithm with a time step of 1.0 fs. The initial configurations were first subjected to energy minimization and subsequently melted in the canonical (*NVT*) and isothermal–isobaric (*NPT*) ensembles at 4000 K and zero pressure (where applicable) for 100 ps, thereby eliminating any residual memory of the initial configuration. The molten phases were then gradually cooled to 300 K under zero pressure at a cooling rate of 1 K/ps to form the glass structure. Finally, the resulting configurations were equilibrated at 300 K in the *NPT* ensemble for 1.0 ns to obtain the final glass structure.

The mechanical behavior of the CAS glasses was determined through fracture simulations following the method of Brochard et al.[38], which is based on the energetic formulation of fracture mechanics. The glass structures were first replicated into 4×4×1 supercells. A pre-crack was then introduced at the center of the structure in the form of an ellipsoidal cylinder with a height of $L_x/15$ and a length of $L_y/3$, where $L_x$ and $L_y$ denote the dimensions of the simulation box in the x and y directions, respectively. The pre-cracked structures were equilibrated in the *NPT* ensemble at 300 K and zero pressure for 100 ps, which was sufficient to ensure convergence of the potential energy of the modified structure. Subsequently, uniaxial tensile loading was applied along the x direction in the *NVT* ensemble at a strain rate of $10^9$ s$^{-1}$. The stress–strain curves were obtained by recording the pressure in the x direction and the corresponding strain during the tensile process until fracture. The Young's modulus (*E*) was determined from the slope of the stress–strain curve in the low-strain region ($\varepsilon < 0.05$) using linear regression. The critical energy release rate ($G_c$) under the assumption of elongation along the x-direction was calculated by integrating the stress–strain curve up to the point of fracture:

$$G_c = \frac{L_y L_z}{\Delta A_\infty} \int_{L_{x,0}}^{L_{x,max}} \sigma_x dL_x \qquad (8)$$

where $L$ denotes the dimension of the simulation box in the specified direction, $\Delta A_\infty$ represents the newly formed crack surface area during fracture, and $\sigma_x$ is the stress value. The fracture toughness $K_{Ic}$ was then calculated using the Irwin formula:

$$K_{Ic} = \sqrt{\frac{G_c E}{1-\nu^2}} \qquad (9)$$

where $\nu$ denotes Poisson's ratio, determined from the stiffness matrix elements $C_{11}$ and $C_{12}$ by stepwise deforming the simulated systems, according to: $\nu = C_{12}/(C_{12} + C_{11})$.

The changes in coordination number (CN) and bond switching activities were analyzed during tensile deformation. Bond-switching analysis quantified the fraction of atoms with increased, decreased, swapped, or unchanged CNs relative to the initial, non-strained structure. This was accomplished by comparing the CN of each atom and the identity of its neighboring atoms with those in the initial configuration. Atoms whose CN decreased or increased are here referred to as decreased CN (DC) or increased CN (IC), respectively. Atoms whose neighboring atoms remained unchanged are referred to as unchanged CN. Otherwise, atoms are designated as swapped CN (SC), indicating that the CN remained the same but at least one neighboring atom had been exchanged. The number of bond switching events for each of these mechanisms were calculated as a fraction of atoms undergoing said mechanism out of the initial configuration for each cation (Ca, Al and Si). The accumulated number of bond switching events (ABSE) was calculated as,

$$ABSE = \sum_i (IC_i + DC_i + SC_i) \cdot X_i \qquad (10)$$

where $IC_i$, $DC_i$ and $SC_i$ are the fractions of atoms that have undergone increase in CN, decrease in CN and swapped CN, respectively, for a given cation $i$ ($i$ = Ca, Al, Si). $X_i$ is the mol fraction of the given cation $i$. ABSE was calculated both as a sum for all cations (Ca, Al and Si) and for Si and Al atoms individually.

## 3. Results and Discussion

*3.1 Experimental and simulated glass structures*

The measured compositions of the nine prepared glasses are given in **Table 1**. Small discrepancies exist between the nominal and actual compositions caused by mass loss during melting, but we infer that the differences between individual samples are sufficient to draw conclusions regarding the composition dependence of structure and mechanical properties in both Series I and II. The glasses were successfully synthesized as they exhibit non-crystallinity (see XRD data in Supporting Figure S1) and glass transitions (see DSC curves in Supporting Figure S2). The respective $T_g$ values, densities, and calculated oxygen packing density can also be found in **Table 1.**

**Table 1.** Analyzed chemical compositions, fraction of oxygens present as triclusters ($X_{O^3}$) obtained from MD simulations, experimental and simulated densities ($\rho$ and $\rho_{sim}$, respectively), glass transition temperature ($T_g$), oxygen packing density ($V_O$), and $R$ composition parameter ($R$ = Al/(Al+Si)) for the two prepared glass series. Sample IDs reflect the compositions as C$x$A$y$S$z$, where $x$ is the amount of calcia, $y$ the amount of alumina, and $z$ the amount silica (in mol%). The standard deviations are ⩽ 0.003 $g/cm^3$ for the experimental densities.

| Sample ID | Composition (mol%) | | | $X_{O^3}$ | $\rho$ | $\rho_{sim}$ | $T_g$ | $V_O$ | $R$ |
|---|---|---|---|---|---|---|---|---|---|
| | CaO | Al$_2$O$_3$ | SiO$_2$ | | g/cm$^3$ | g/cm$^3$ | °C | | |
| Series I | | | | | | | | | |
| C19A18S64 | 18.6 | 17.6 | 63.8 | 0.05 | 2.56 | 2.55 | 856 | 0.48 | 0.35 |
| C25A23S52 | 25.1 | 22.8 | 52.1 | 0.06 | 2.69 | 2.67 | 848 | 0.50 | 0.47 |
| C30A29S41 | 30.3 | 29.2 | 40.5 | 0.08 | 2.75 | 2.77 | 853 | 0.52 | 0.59 |
| C36A33S31 | 35.9 | 33.2 | 30.9 | 0.10 | 2.80 | 2.79 | 855 | 0.52 | 0.68 |
| C40A40S20 | 40.1 | 39.9 | 20.0 | 0.13 | 2.84 | 2.93 | 866 | 0.54 | 0.80 |
| Series II | | | | | | | | | |
| C26A33S41 | 25.6 | 33.2 | 41.2 | 0.14 | 2.75 | 2.75 | 857 | 0.52 | 0.62 |
| C30A29S41 | 30.3 | 29.2 | 40.5 | 0.08 | 2.75 | 2.77 | 853 | 0.52 | 0.59 |
| C36A24S40 | 35.5 | 24.3 | 40.2 | 0.04 | 2.77 | 2.84 | 832 | 0.50 | 0.55 |
| C40A19S41 | 40.2 | 19.3 | 40.5 | 0.02 | 2.82 | 2.80 | 814 | 0.50 | 0.49 |
| C45A15S40 | 45.3 | 14.6 | 40.1 | 0.01 | 2.86 | 2.85 | 802 | 0.48 | 0.42 |

Correlations between the simulated and experimental densities and Young's moduli can be found in Supporting Figure S3. Density features a better agreement between simulated and experimental values for Series I than Series II, but overall, the simulations capture the correct trend in density variation for both series. That is, the densities increase with decreasing silica content in Series I and with increasing Ca/Al ratio in Series II. This is in contrast to the oxygen packing density ($V_O$), which increases with the alumina content for both series. The change in density in both series thus appear to be mainly related to the addition of calcia, as Ca has a higher atomic mass relatively to Al and Si and the fact that modifier ions occupy the interstitial spaces in the network. We also find that $T_g$ generally increases with the alumina content, with the exception of the C19A18S64 sample in Series I. This sample has the lowest amount of alumina in that series but also the lowest amount of modifier, i.e., the minimum $T_g$ in this series is found for sample C25A23S52 due to its relatively small amount of calcia compared to the amount of silica.

The experimental reduced pair correlation functions, $G(r)$, as obtained from X-ray total scattering measurements, are depicted in **Figure 3a** and **3b** for Series I and II glasses, respectively. These functions are plotted for $r$ up to 12 Å in Supporting Figure S4.[39] The peak position at ~1.61-1.75 Å represents Si-O bonds (low-$r$ side) and Al-O bonds (high-$r$ side). These peaks are merged but are assigned based on the work of Petkov et al.[40] and Hennet et al.[40] In Series I, this peak shifts towards longer distance and decreases slightly in intensity with increasing alumina content. In Series II, where the silica content is constant, there is only a small shift to longer distance but a drastic increase in intensity with increasing alumina content. The second peak at ~2.37 Å represents Ca-O bonds and it increases in intensity with increasing calcium content for both series.

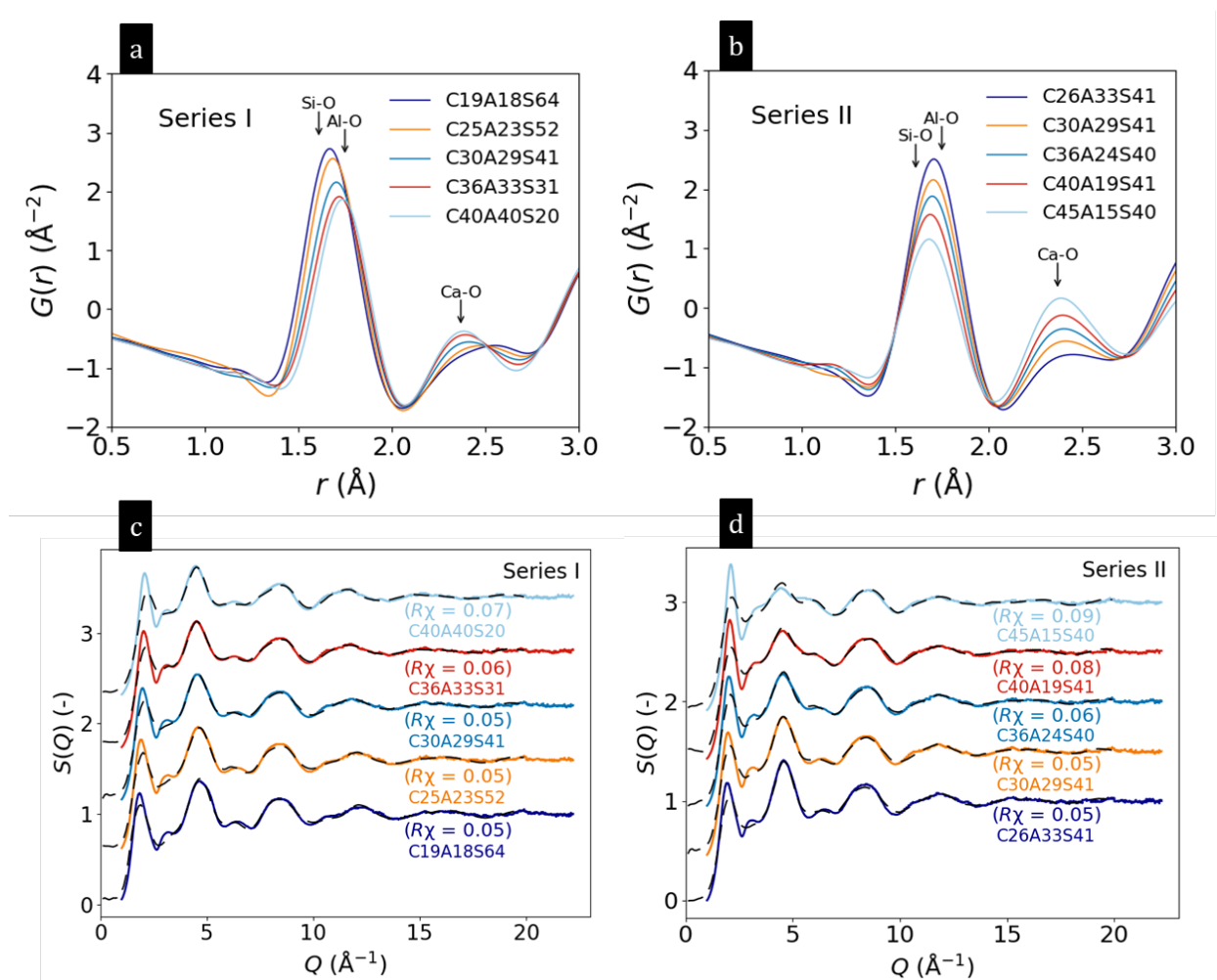

**Figure 3.** X-ray total scattering results. (a-b) Experimental reduced pair correlation functions, $G(r)$, of (a) Series I and (b) Series II glasses. All $G(r)$ data was obtained using Lorch like modification and $Q_{max}$ = 22.2 Å$^{-1}$. (c-d) Experimental (full line) and MD-simulated (dashed line) X-ray structure factors, $S(Q)$, of (c) Series I and (d) Series II glasses. The different compositions are plotted with an offset of 0.5 on the y-axis in panels (c) and (d).

Next, we compare the experimental structure data with those of the MD-simulated glasses by considering the X-ray structure factor, $S(Q)$. Results are shown in **Figures 3c** and **3d** for Series I and II glasses, respectively, with simulated structures depicted as dashed lines and experimental structures depicted as solid lines. To quantify the agreement between the simulated and experimental structures, we calculate the $R\chi$ factor,[20,41]

$$R_\chi = \left( \frac{\sum_i [S_{\exp}(Q_i) - S_{\text{sim}}(Q_i)]^2}{\sum_i S_{\exp}^2(Q_i)} \right)^{\frac{1}{2}} \tag{11}$$

$R_\chi$ is calculated in the $Q$ range from 1 to 8 Å$^{-1}$ [20,41]. In general, we find good agreement between the simulated and experimental $S(Q)$, with the highest $R_\chi$ value found for C45A15S40 to be 0.09 which we deem to be acceptable within this $Q$ range[20,41]. The discrepancies between the experimental and simulated structures become increasingly prominent with increasing calcia content.[42] This suggests that the MD-simulations are less effective in describing the chemical environment of the calcium cations than the network forming ions. In any case, the peak positions in the simulated and experimental data are generally in good agreement. Based on this and the generally well-replicated densities, we infer that the MD-simulated glass structures can be used for analyzing their mechanical properties through fracture simulations.

*3.2 Elasticity, hardness, and crack resistance*

All the experimentally measured elastic moduli are given in Supporting Table S1, while both the measured and simulated Young's modulus can be found in **Table 2** along with all the other measured mechanical properties and the ABSE that have been derived from the fracture simulations. Vickers hardness increases with increasing concentration of alumina for both Series I and II glasses (**Figure 4a**), in agreement with previous findings[25–27]. The work of Pönitzsch et al.[27] suggests that hardness steeply increases with alumina content in the peraluminous region, but this is not observed here, as the investigated composition in the peraluminous region (C26A33S41, $R$=0.62) has similar hardness as the tectosilicate composition in the same series. In addition, we note that the hardness increases in parallel with Young's modulus when silica and calcia are substituted for alumina, except for the above-mentioned measurement in the peraluminous region. NBOs make the glass network less constrained, resulting in a softer material that is more prone to deformation[16]. In agreement with this, increasing calcia content in Series II leads to more NBOs, thus reducing hardness and Young's modulus. Since the increase in calcia content for the tectosilicate glasses in Series I does not lead to more NBOs (as these glasses are ideally fully polymerized), more calcia does not cause the same effect in Series I.

The silica tetrahedral basic units contribute to a larger free volume, and glasses with a larger free volume are more prone to densification through bond angle distortion when stress is applied and consequently they have relatively low hardness and Young's modulus.[42] In this study, we find that the composition with the lowest Young's modulus (C19A18S64, $R$=0.35) has the highest silica content and the lowest hardness of Series I. According to the work of Shan et al.[21], the amount of oxygen triclusters and Al$^{V}$ increase with increasing alumina content in the peraluminous region. The present MD simulations results confirm that the fraction of oxygen triclusters ($X_{O^3}$) increases with increasing amount of alumina (**Table 1**). Both oxygen triclusters and $Al^V$ inhibit further indentation-induced densification as they result in higher network connectivity, which causes the hardness to increase[8,21]. Here, we find that the increase in hardness matches the trend in increasing $V_O$ with increasing alumina content (**Table 1**). For Series II, $V_O$ increases when Ca is replaced with Al, which is a consequence of the amount of oxygen in the structure increasing with the concentration of alumina, as alumina contributes with 1½ oxygen atom per Al and calcia contributes with 1 oxygen atom per Ca. Furthermore, the mean volume that one oxygen occupies (**Eqs. 3**) is calculated for the oxygens in the network only (i.e., Si-O and Al-O). For Series I, $V_O$ increases when Si is replaced by Ca and Al. This is a consequence of the Si basic units contributing to a larger free volumes, increasing the propensity for indentation-induced densification, which can be observed from the side length recovery ratio ($L_{SR}$) data in **Figure 4b.**

**Table 2.** Experimental Poisson's ratio ($v$), experimental and simulated Young's modulus ($E$ and $E_{sim}$), crack resistance ($CR$), Vickers hardness ($H_V$), side length recovery ratio ($L_{SR}$), fracture toughness values ($K_{Ic}$): experimental, calculated by Rouxel's model (mod), and MD simulated (sim), and accumulated number of bond switching events (ABSE) for Si and for Al (as obtained from MD simulations).

| Sample ID | $v$ | $E$ GPa | $E_{sim}$ GPa | $CR$ N | $H_V$ GPa | $L_{SR}$ | $K_{Ic}$ MPa·m$^{1/2}$ | $K_{Ic,mod}$ MPa·m$^{1/2}$ | $K_{Ic,sim}$ MPa·m$^{1/2}$ | ABSE for Si | ABSE for Al |
|---|---|---|---|---|---|---|---|---|---|---|---|
| Series I | | | | | | | | | | | |
| C19A18S64 | 0.255 | 88 | 91.4 ± 0.7 | 4.4 | 7.0 ± 0.2 | 0.22 ± 0.04 | 0.86 ± 0.09 | 0.81 | 0.76 ± 0.11 | 0.33·10$^{-2}$ | 1.30·10$^{-2}$ |
| C25A23S52 | 0.271 | 95 | 92.6 ± 0.6 | 3.2 | 7.0 ± 0.1 | 0.20 ± 0.03 | 0.85 ± 0.03 | 0.84 | 0.67 ± 0.02 | 0.13·10$^{-2}$ | 1.44·10$^{-2}$ |
| C30A29S41 | 0.276 | 99 | 95.1 ± 0.4 | 6.5 | 7.2 ± 0.1 | 0.17 ± 0.05 | 0.91 ± 0.03 | 0.85 | 0.67 ± 0.01 | 0.05·10$^{-2}$ | 1.70·10$^{-2}$ |
| C36A33S31 | 0.282 | 102 | 102.7 ± 0.5 | 5.3 | 7.3 ± 0.2 | 0.10 ± 0.04 | 0.95 ± 0.02 | 0.85 | 0.68 ± 0.02 | 0.05·10$^{-2}$ | 2.00·10$^{-2}$ |
| C40A40S20 | 0.284 | 106 | 110 ± 0.5 | 5.5 | 7.6 ± 0.2 | 0.04 ± 0.03 | 0.94 ± 0.07 | 0.85 | 0.71 ± 0.02 | 0.03·10$^{-2}$ | 2.40·10$^{-2}$ |
| Series II | | | | | | | | | | | |
| C26A33S41 | 0.276 | 102 | 105.2 ± 0.4 | 10.1 | 7.3 ± 0.3 | 0.22 ± 0.03 | 0.97 ± 0.04 | 0.86 | 0.75 ± 0.02 | 0.14·10$^{-2}$ | 2.51·10$^{-2}$ |
| C30A29S41 | 0.276 | 99 | 95.1 ± 0.4 | 6.5 | 7.2 ± 0.1 | 0.17 ± 0.05 | 0.91 ± 0.03 | 0.85 | 0.67 ± 0.01 | 0.05·10$^{-2}$ | 1.70·10$^{-2}$ |
| C36A24S40 | 0.279 | 97 | 96.0 ± 1.8 | 3.0 | 7.0 ± 0.1 | 0.09 ± 0.04 | 0.91 ± 0.04 | 0.84 | 0.66 ± 0.00 | 0.07·10$^{-2}$ | 1.40·10$^{-2}$ |
| C40A19S41 | 0.280 | 96 | 97.5 ± 1.3 | 2.9 | 6.7 ± 0.2 | 0.02 ± 0.02 | 0.94 ± 0.06 | 0.84 | 0.65 ± 0.02 | 0.09·10$^{-2}$ | 1.38·10$^{-2}$ |
| C45A15S40 | 0.283 | 95 | 90.3 ± 0.7 | 2.5 | 6.5 ± 0.1 | 0.01 ± 0.02 | 0.87 ± 0.02 | 0.84 | 0.61 ± 0.02 | 0.10·10$^{-2}$ | 0.97·10$^{-2}$ |

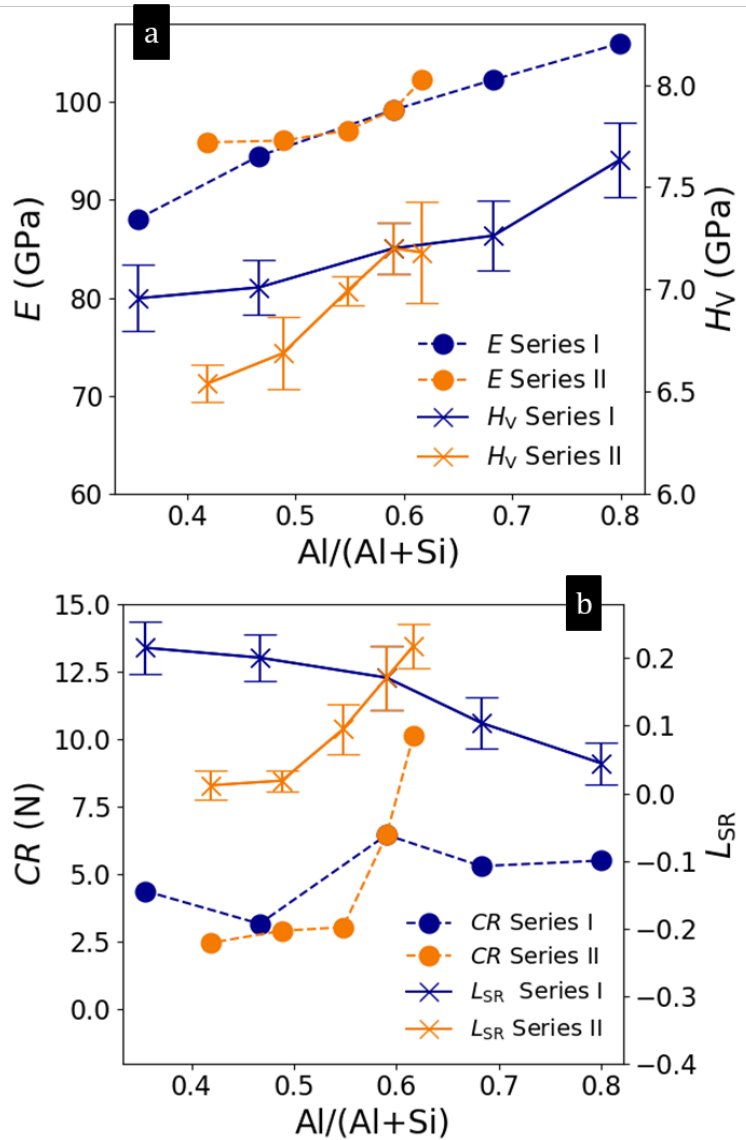

**Figure 4.** (a) Composition dependence of Vickers hardness and Young's modulus. (b) Composition dependence of crack resistance and side length recovery ratio (unitless).

The side length recovery ratio ($L_{SR}$) data show that the high-silica glasses densify more during indentation than high-alumina glasses, with the smallest amount of densification observed in glasses with high content of calcia (**Figure 4b**). This agrees with the fact that peralkaline glasses should be more prone to shear flow because of more NBOs in their structure[6,16]. Deformation through shear flow is not recovered during the applied $0.9T_g$ heat treatment, while the densified volume does recover. Shan et al.[21] have found that crack resistance increases with the alumina content, which agrees with the trend found herein for Series II (**Figure 4b**). However, for Series I, the crack resistance does not show a monotonic increase with increasing alumina content but rather an optimum at 41 mol% silica ($R=0.59$). The crack resistance is expected to increase as the fraction of $Al^V$ increases, since the bonds of $Al^V$ species are more ionic than $Al^{IV}$, making $Al^V$ more prone to bond switching[21]. This is in agreement with the simulated ABSE for the network ions (Si and Al) in the glasses (**Table 2** and Supporting Figure S5a), at least for Series II.

Furthermore, there is a clear positive correlation between $X_{O^3}$ and ABSE for Al (Supporting Figure S5b). An increase in crack resistance as a function of the ABSE for Al+Si can only be found for silica content up to 41 mol% ($R$=0.59) in Series I, while the ABSE continues to increase with decreasing amount of silica (Supporting Figure S5a). A larger amount of modifier results in less free volume and thus less densification, as seen from the significant decrease in $L_{SR}$ when the calcia content increases. The lower amount of densification with increasing calcia content could be the explanation for the decrease in crack resistance at silica contents below 41 mol% ($R$=0.59) (Supporting Figure S5c).

Correlation plots between the mechanical properties ($CR$ and $K_{Ic}$) and the ABSE with contribution from all cations can be found in Supporting Figure S6. We note, however, that it is challenging to analyze the bond switching events for the ionic Ca-O bonds compared to the iono-covalent Al-O and Si-O bonds of the network because of the lack of a clear cut-off distance for the Ca-O bond. To avoid an unreliable correlation between the mechanical properties and the bond switching events, the bond switching events for Ca are therefore omitted. In detail, the ABSE presented in **Table 2** are calculated as the ABSE for Si and Al individually but are presented as a sum of the two in the figures.

*3.3 Fracture toughness*

In both series of glasses, higher alumina content increases the experimentally measured fracture toughness $K_{Ic}$ (**Figure 5a**). We note that the standard deviations of $K_{Ic}$ are relatively small (less than ±0.04 MPa m$^{0.5}$), except for C19A18S64, C40A40S20, and C40A19S41 glasses that also seem to deviate the most from the observed trends for each series. To understand these $K_{Ic}$ results, we have performed MD fracture simulations (yielding $K_{Ic,sim}$) and calculated theoretical fracture toughness values ($K_{Ic,mod}$), as shown in **Table 2**. The latter is done using Rouxel's model[10,15,17], as $G_c$ is replaced by $2\gamma$ in **Eqs 9.**

$$\gamma = \frac{1}{2}\left(\frac{\rho}{M_0}\right)^{2/3} N_A^{-1/3} \sum_i x_i n_i U_{oi} \qquad (12)$$

Here, $\gamma$ is the fracture surface energy, $N_A$ is Avogadro's number, $x_i$ is the fraction of each cation ($i$=Ca, Al, Si), $n_i$ is the number of broken bonds per cation unit and $U_{oi}$ is the diatomic bond energy. The diatomic bond energies are 383.3, 501.9, and 799.6 kJ/mol for Ca-O, Al-O, and Si-O, respectively[43].

The $K_{Ic,mod}$ values in **Table 2** are calculated assuming that each type of cation contributes equally to the crack path. Rouxel[17] suggests that when the alkaline-earth content exceeds more than half of the silica content, silica should be excluded from the calculations as the crack will always follow the path of least resistance. As Ca and Al have lower bond energies than Si, these create the path of least resistance, i.e., the crack will not propagate through the Si bonds, when the content of silica is low enough. As only one composition (C19A18S64) has a calcium content below half of the silica content, calculations of $K_{Ic,mod}$ excluding the contribution of the Si bonds can be found in Supporting Table S2. The $K_{Ic,mod}$ values with full contributions are, however, still lower than the experimental $K_{Ic}$ values, suggesting that the path of least resistance does not seem to fully apply for the CAS glasses. Correlation plots between the different fracture toughness values can be found in Supporting Figures S7.

The observed composition dependence of $K_{Ic,sim}$ captures the experimental trend relatively well for Series II, but not for Series I (**Figure 5a**). In general, the $K_{Ic,sim}$ values are lower than the experimental $K_{Ic}$, which can be due to differences in system size and strain rate[44,45]. For example, as the simulations are made on the nanoscale, it is not possible to capture microscale relaxation, and the higher strain rate in simulations give less time for the system to undergo plastic relaxation, bond switching, and stress redistribution in the structure. The model of Rouxel generally offers good predictions of the experimental fracture toughness values for both series, but similar to the MD simulations results, the values are generally underestimated (**Figure 5a**). It has previously been reported that this model underestimates fracture toughness when atoms with a high propensity for bond switching, such as B or Al, are present.[14,17] This might be because the model assumes perfectly brittle crack growth, i.e., any nanoscale plastic deformation is neglected (e.g., crack tip bluntness caused by corrosion or crack tip relaxation, which can occur in glasses with nanoscale ductility[17]). In the present case, the model may thus underestimate fracture toughness as Al should increase the nanoscale ductility in form of bond switching.[3,14,15] To this end, Supporting Figure S8 shows the difference between the $K_{Ic}$ and the $K_{Ic,mod}$ as a function of the ABSE for the network cations. If we do not consider the $K_{Ic}$ values with errors above 0.05 MPa·m$^{1/2}$, we find that the $K_{Ic}$-$K_{Ic,mod}$ difference is positively correlated with the number of bond switching events. This is indicates that the neglected ductility of Rouxel's model is connected to the bond switching events.

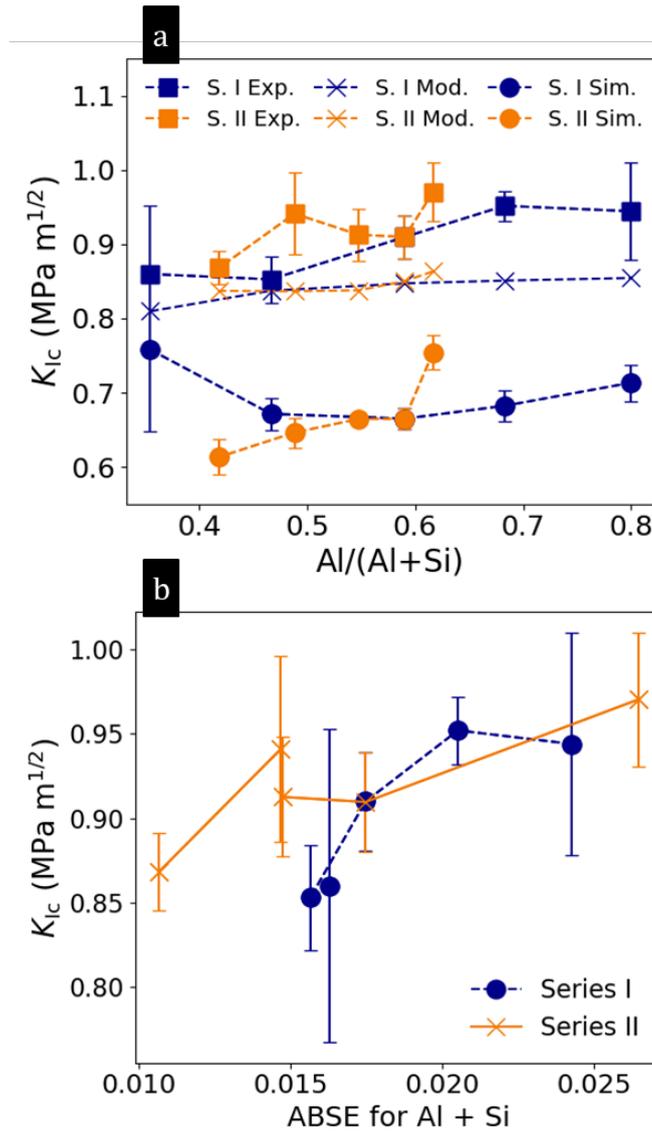

**Figure 5.** (a) Composition dependence of fracture toughness: experimental, MD-simulated and calculated according to Rouxel's model. (b) Dependence of experimental fracture toughness on the accumulated number of bond switching events (ABSE) for Al and Si atoms as determined from MD simulations.

For comparison, Gross et al.[11] measured the fracture toughness of C21A20S59 and C25A24S50 glasses to be 0.893 and 0.898 MPa m$^{0.5}$, respectively, using the chevron notched short bar method. Comparing these values to those of similar glass compositions in this study (C19A18S64 and C25A23S52), we find that the values measured by Gross et al.[11] are even higher than those measured herein of 0.860 and 0.853 MPa m$^{0.5}$, respectively. This further supports that the Rouxel model indeed underestimates the fracture toughness values for CAS glasses, although the overall compositional trends are well captured.

We find that number of MD-simulated bond switching events for Al and Si correlates positively with the experimental fracture toughness for both series (**Figure 5b**). Considering the different bond switching mechanisms, i.e., decrease in coordination number (DC), increase in coordination number (IC), and swapped coordination number (SC), we find that SC is the most dominant mechanism for Al-O (Supporting Figure S9b). The same applies for Si-O (Supporting Figure S9c), but the number of bond switching events for Si is significantly smaller than that of Al-O (**Table 2**). The SC for Al is mostly attributed to the Al$^\text{V}$ in the structure as these are most prone to bond switching.[21] Notably, the bond switching for Ca-O does not seem to correlate with the fracture toughness for Series II. This could be because the dominant type of bond switching for Ca-O is DC (Supporting Figure S9a), whereas the least dominant type is SC. We note that an increase in network connectivity induced by hot compression has been found to increase both fracture toughness and fraction of bond switching events.[14] This is in agreement with the increase in connectivity with increasing alumina content, as observed from the increase in $V_\text{O}$ and the fact that species such as Al$^\text{V}$ and oxygen triclusters contribute to a higher structural connectivity[21,27]. Furthermore, To et al.[14] found that the prediction of fracture toughness could be improved by including more bonds in the pathway when calculating the fracture surface energy. This again suggests that the exclusion of bond switching in Rouxel's model is what causes the underestimation of the fracture toughness.

## 4. Conclusions

The mechanical properties, particularly fracture toughness, of different calcium aluminosilicate glasses have been evaluated to understand how they correlate with the underlying glass structure and structural changes during crack growth. In detail, two series of glasses have been investigated: tectosilicate glasses with 20-60 mol% silica and glasses with constant silica content (41 mol%) but varying Al$_2$O$_3$/CaO ratio from 0.3 to 1.3. We have also estimated Al and Si bond switching events through MD simulations, where the mechanisms are increased coordination number (IC), decreased coordination number (DC) and finally swapped coordination (SC). The experimentally measured hardness and Young's modulus increase monotonically with the alumina content, and the increase in alumina content also increases the crack resistance at silica contents at or above 41 mol%, with a steep increase in the peraluminous region. For silica contents above 41 mol% the crack resistance appears to be dependent on the number of Al and Si bond switching events, with limited densification upon indentation causing the lower crack resistance for silica contents below 41 mol%. The experimentally determined fracture toughness also increases with alumina content for both series of glasses, and similarly, the fracture toughness can be connected to the number of bond switching events, especially the swapped coordination. As Al-O contributes with a significantly higher degree of bond switching than Si-O, alumina is indeed crucial for the mechanical properties of the investigated glasses. Overall, we therefore conclude that the most crack resistant and toughest glasses are found in the peraluminous region.

## Acknowledgements

We thank Marco Holzer (Federal Institute of Materials Research and Testing) for helpful discussions, and Randall E. Youngman and Abigail Austin (Corning Incorporated) for providing the chemical composition analyses. This work was supported by the European Union (ERC, NewGLASS, 101044664). Views and opinions expressed are, however, those of the authors only and do not necessarily reflect those of the European Union or the European Research Council. Neither the European Union nor the granting authority can be held responsible for them. We also acknowledge the support for high-energy synchrotron X-ray total scattering experiments at the 3W1 beam line of Beijing Synchrotron Radiation Facility (BSRF), China.

# Supporting Information

*for*

# Connecting bond switching to fracture toughness of calcium aluminosilicate glasses

Sidsel Mulvad Johansen[1], Tao Du[1,2], Johan F. S. Christensen[1], Anders K. R. Christensen[1], Xuan Ge[1,3], Theany To[4,5], Lars R. Jensen[6], Morten M. Smedskjaer[1,*]

[1]Department of Chemistry and Bioscience, Aalborg University, 9220 Aalborg East, Denmark

[2]Department of Applied Physics, The Hong Kong Polytechnic University, Kowloon, Hong Kong, 999077, China

[3]Shanghai Key Laboratory of Materials Laser Processing and Modification, School of Materials Science and Engineering, Shanghai Jiao Tong University, 200240 Shanghai, China

[4]Univ Rennes, CNRS, IPR (Institut de Physique de Rennes) - UMR 6251, Rennes, France

[5]Nantes Université, Ecole Centrale Nantes, CNRS, GeM, UMR 6183, Nantes F-44000, France

[6]Department of Materials and Production, Aalborg University, 9220 Aalborg East, Denmark

* Corresponding author. E-mail: mos@bio.aau.dk

**Table S1.** Experimental elastic moduli data for all investigated CAS glasses.

| Sample ID | $E$ GPa | $G$ GPa | $B$ GPa | $v$ |
|---|---|---|---|---|
| Series I | | | | |
| C19A18S64 | 88.1 | 35.1 | 60.0 | 0.255 |
| C25A23S52 | 94.5 | 37.2 | 68.7 | 0.271 |
| C30A29S41 | 99.2 | 38.9 | 73.9 | 0.276 |
| C36A33S31 | 102 | 39.9 | 78.1 | 0.282 |
| C40A40S20 | 106 | 41.3 | 81.7 | 0.284 |
| Series II | | | | |
| C26A33S41 | 102 | 40.1 | 76.2 | 0.276 |
| C30A29S41 | 99.2 | 38.9 | 73.9 | 0.276 |
| C36A24S40 | 97.1 | 38.0 | 73.2 | 0.279 |
| C40A19S41 | 96.1 | 37.5 | 72.7 | 0.280 |
| C45A15S40 | 95.9 | 37.4 | 73.5 | 0.283 |

**Table S2.** Fracture toughness ($K_{Ic}$) and fracture surface energies ($\gamma$) as calculated using Rouxel's model based on two different assumptions. One model is calculated with the assumption that all cations contribute with one broken bond, whereas the second model excludes contributions from Si. Experimental values are included for comparison.

| Sample ID | $K_{Ic}$ (MPa·m$^{1/2}$) | | | $\gamma$ (J/m$^2$) | | |
|---|---|---|---|---|---|---|
| | All cations | Ca and Al | Experimental | All ions | Ca and Al | Experimental |
| Series I | | | | | | |
| C19A18S64 | 0.81 | 0.46 | 0.86 ± 0.09 | 3.48 | 1.14 | 3.92 |
| C25A23S52 | 0.84 | 0.55 | 0.85 ± 0.03 | 3.44 | 1.51 | 3.56 |
| C30A29S41 | 0.85 | 0.63 | 0.91 ± 0.03 | 3.34 | 1.86 | 3.85 |
| C36A33S31 | 0.85 | 0.69 | 0.95 ± 0.02 | 3.26 | 2.14 | 4.08 |
| C40A40S20 | 0.85 | 0.75 | 0.94 ± 0.07 | 3.17 | 2.46 | 3.87 |
| Series II | | | | | | |
| C26A33S41 | 0.86 | 0.65 | 0.97 ± 0.04 | 3.37 | 1.91 | 4.25 |
| C30A29S41 | 0.85 | 0.63 | 0.91 ± 0.03 | 3.34 | 1.86 | 3.85 |
| C36A24S40 | 0.84 | 0.62 | 0.91 ± 0.04 | 3.33 | 1.81 | 3.96 |
| C40A19S41 | 0.84 | 0.60 | 0.94 ± 0.06 | 3.36 | 1.74 | 4.25 |
| C45A15S40 | 0.84 | 0.59 | 0.87 ± 0.02 | 3.36 | 1.68 | 3.62 |

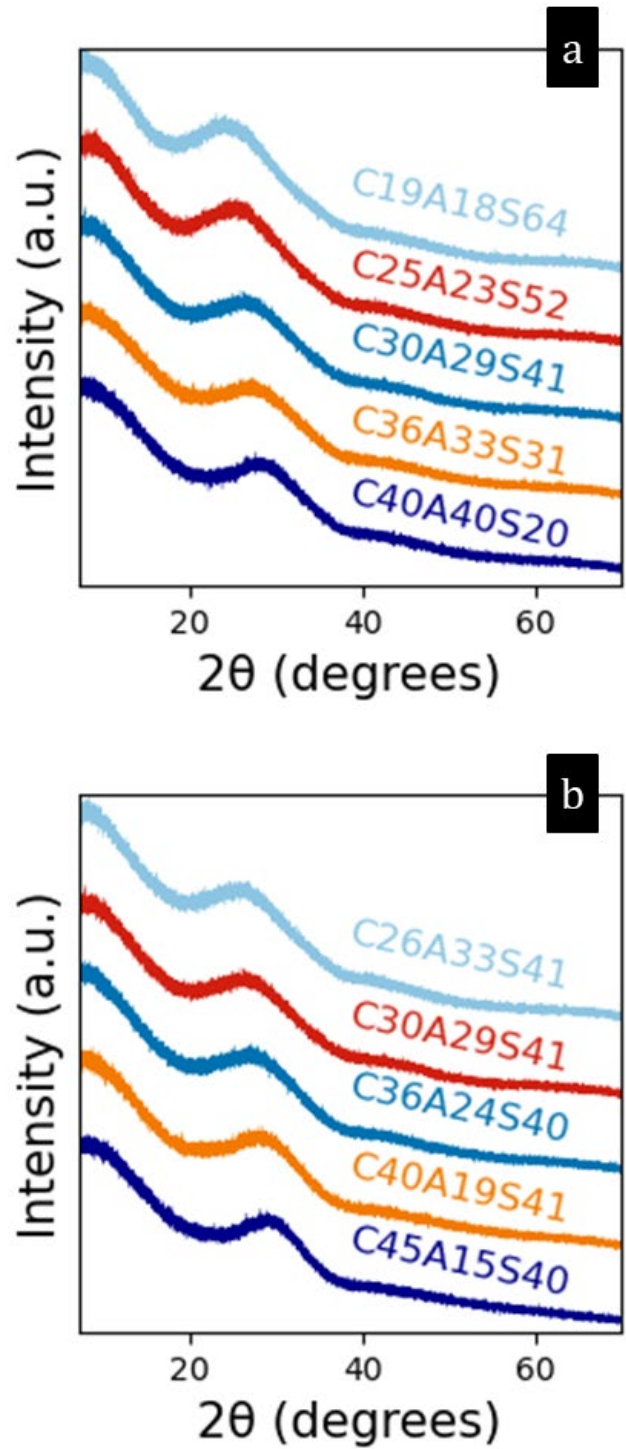

**Figure S1.** XRD diffractograms for (a) Series I and (b) Series II glasses, confirming that all samples are non-crystalline.

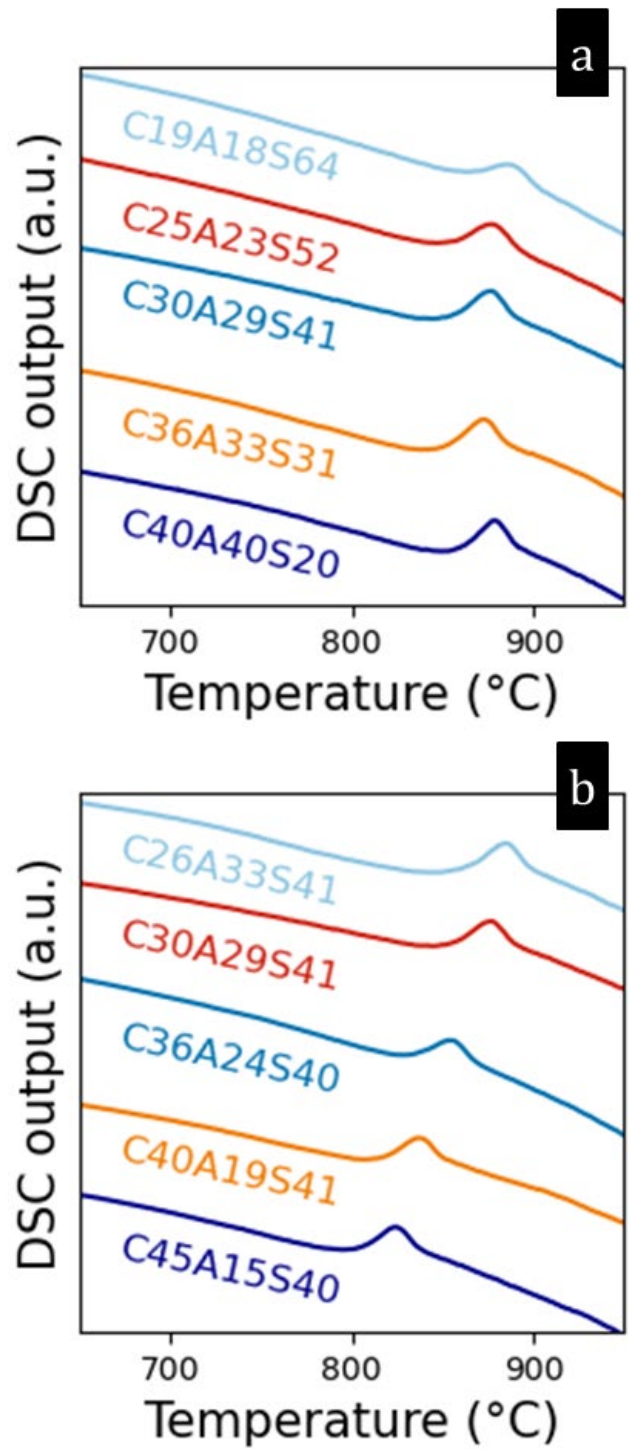

**Figure S2.** DSC heating curves for (a) Series I and (b) Series II glasses.

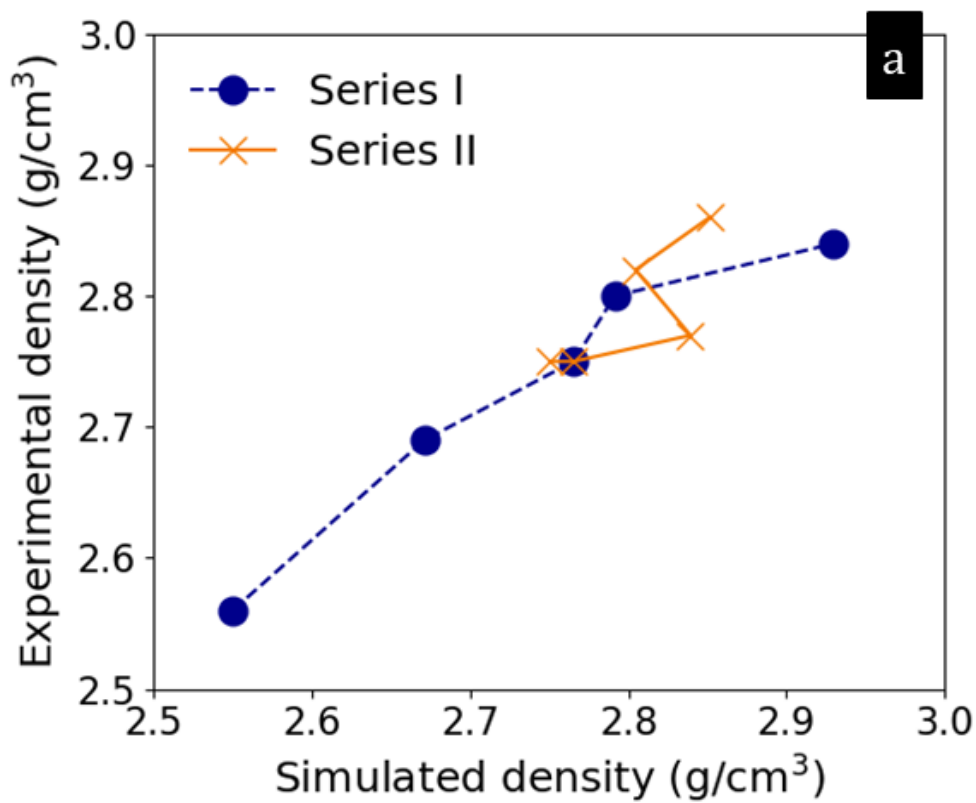

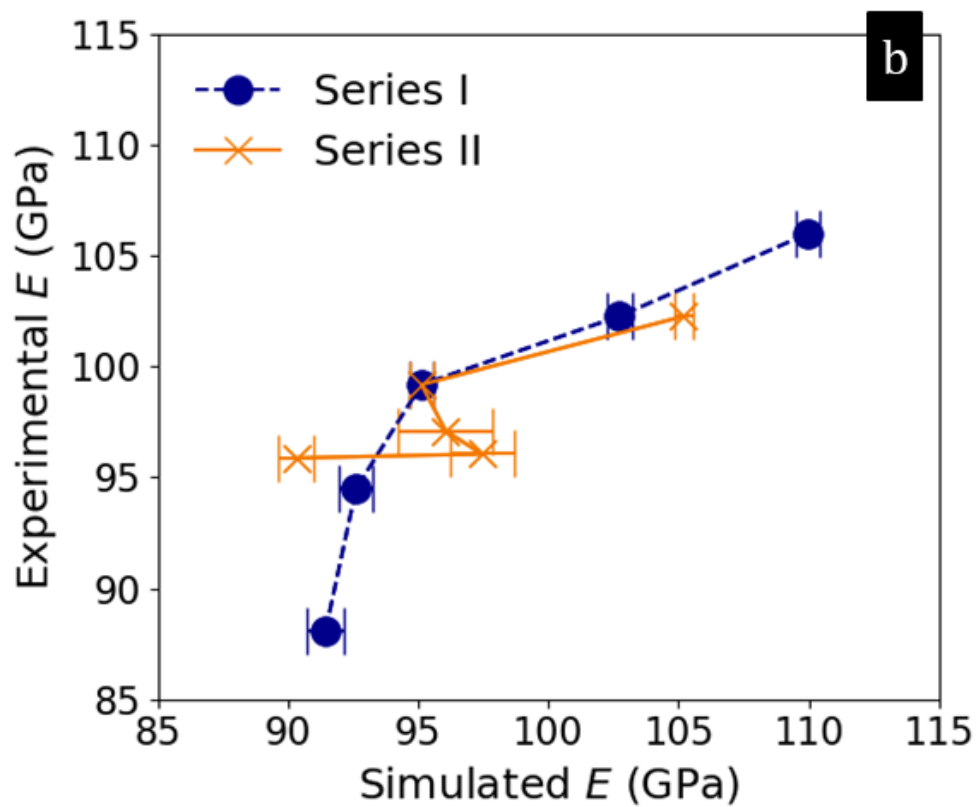

**Figure S3.** (a) Relationship between experimental density and MD-simulated density. (b) Correlation between experimental Young's modulus and MD-simulated Young's modulus.

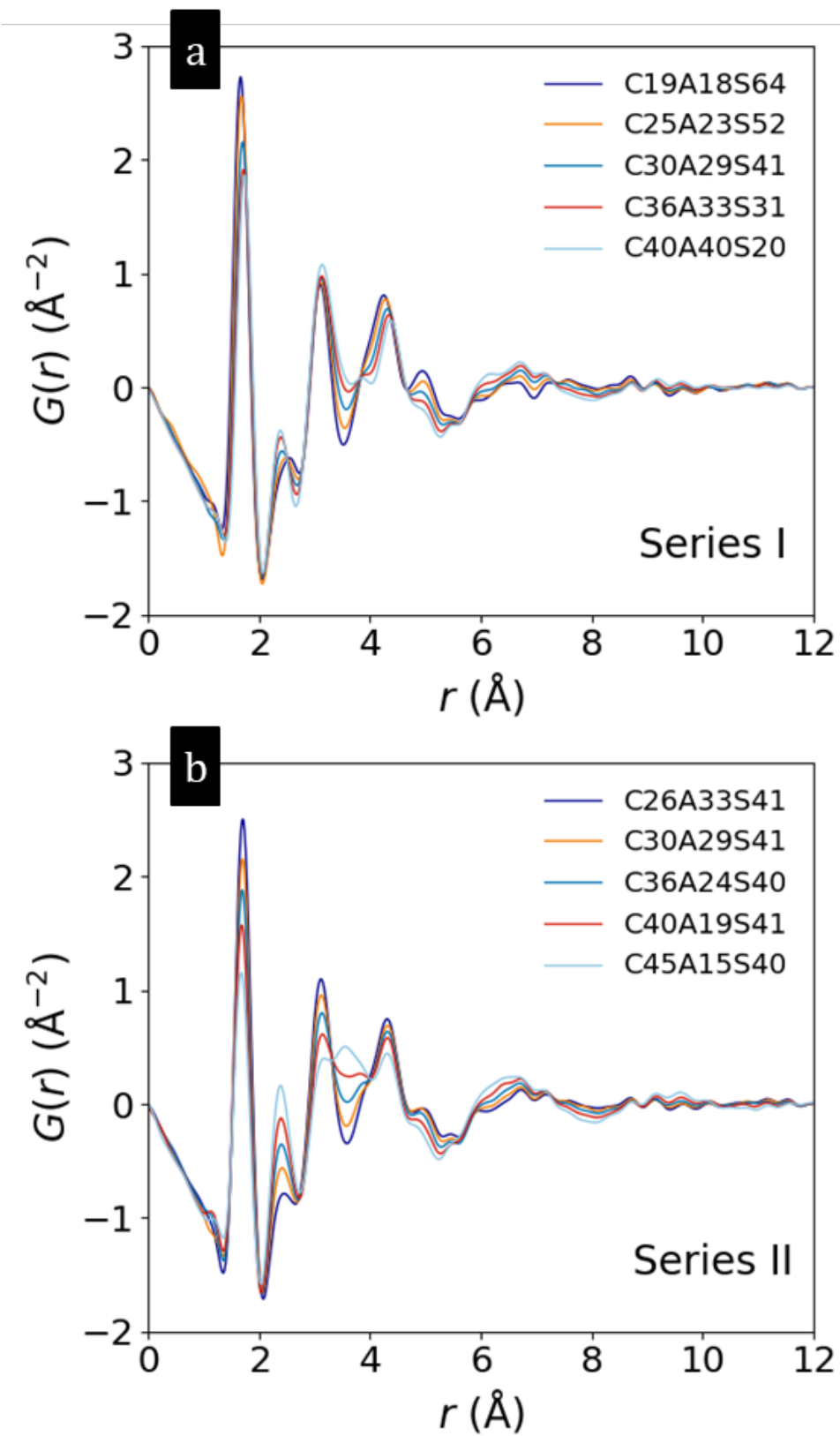

**Figure S4**. X-ray total scattering results. Experimental reduced pair correlation functions of (a) Series I and (b) Series II glasses. All $G(r)$ data was obtained using Lorch like modification and $Q_{max}$ = 22.2 Å$^{-1}$.

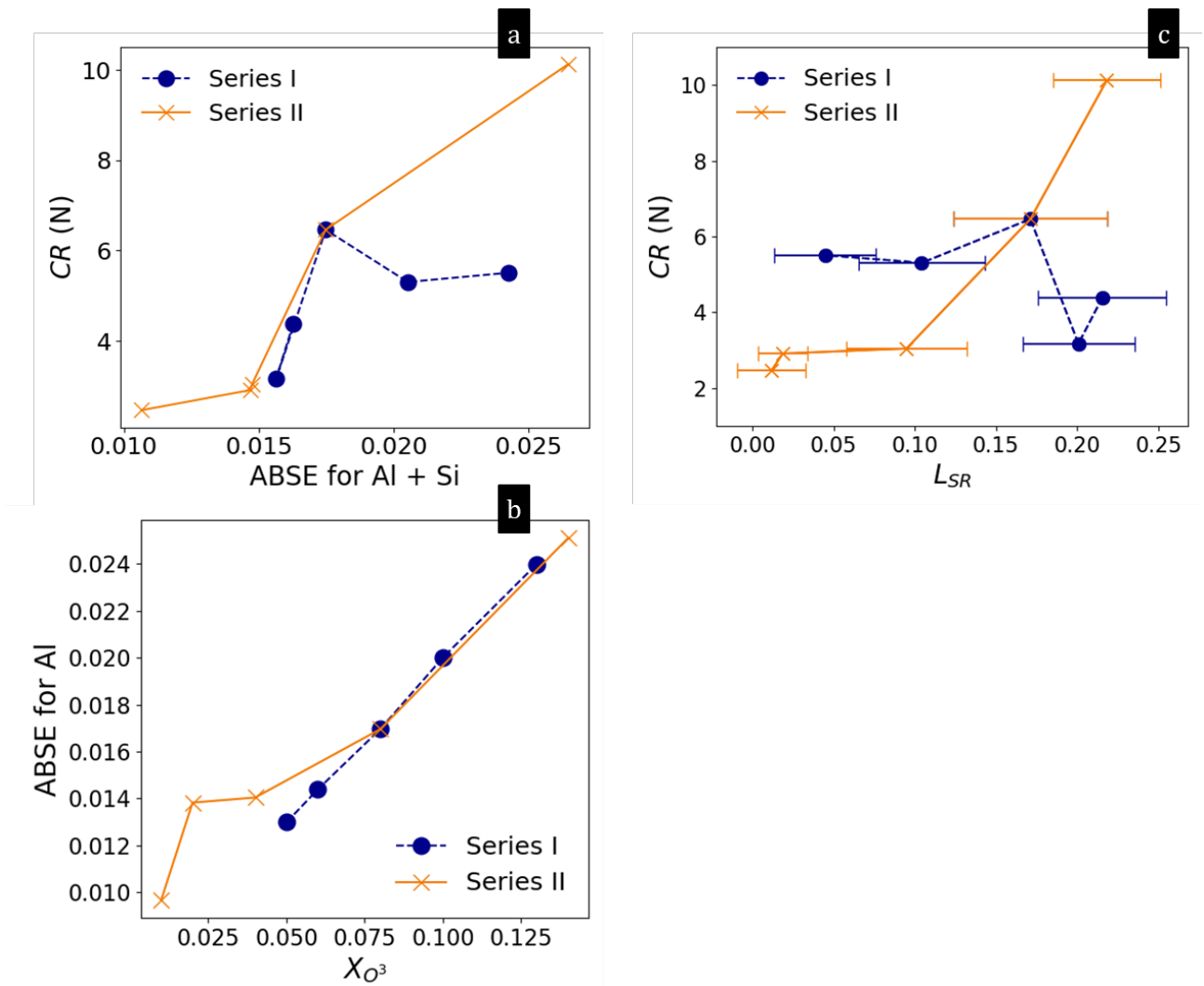

**Figure S5**. (a) Crack resistance (*CR*) as a function of the accumulated number of bond switching events (ABSE) for both Al and Si atoms as obtained from MD simulations. (b) ABSE for Al as a function of the fraction of oxygen triclusters ($X_{O^3}$) as obtained from MD simulations. (c) *CR* as a function of the side length recovery ratio ($L_{SR}$).

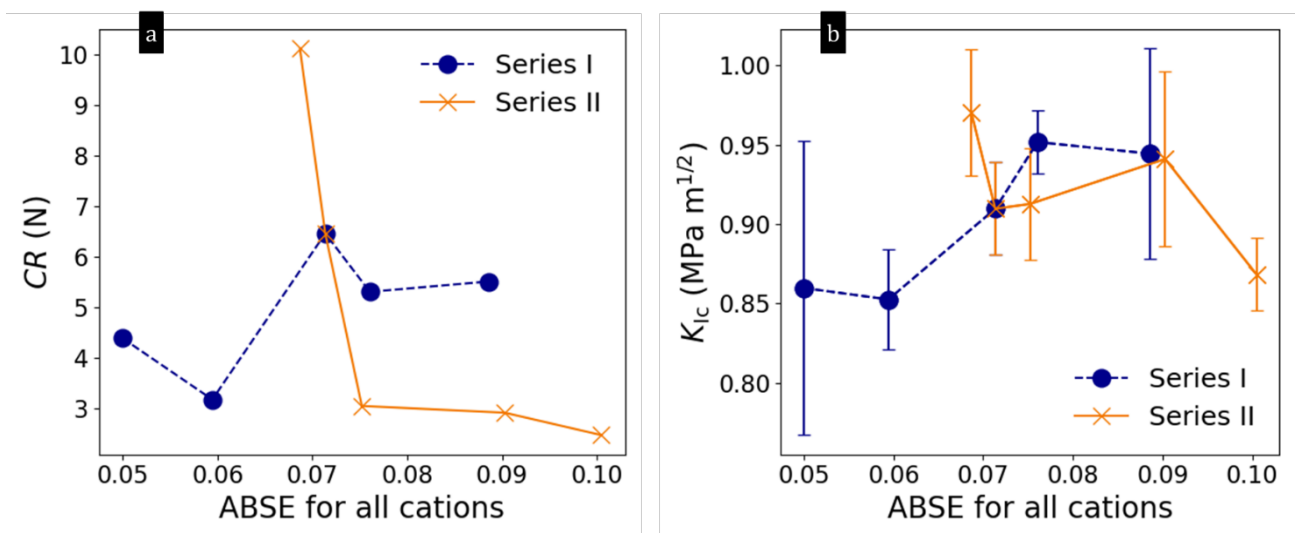

**Figure S6.** Relationship between MD-simulated accumulated number of bond switching events (ABSE) for all cations and (a) crack resistance ($CR$) and (b) fracture toughness ($K_{Ic}$).

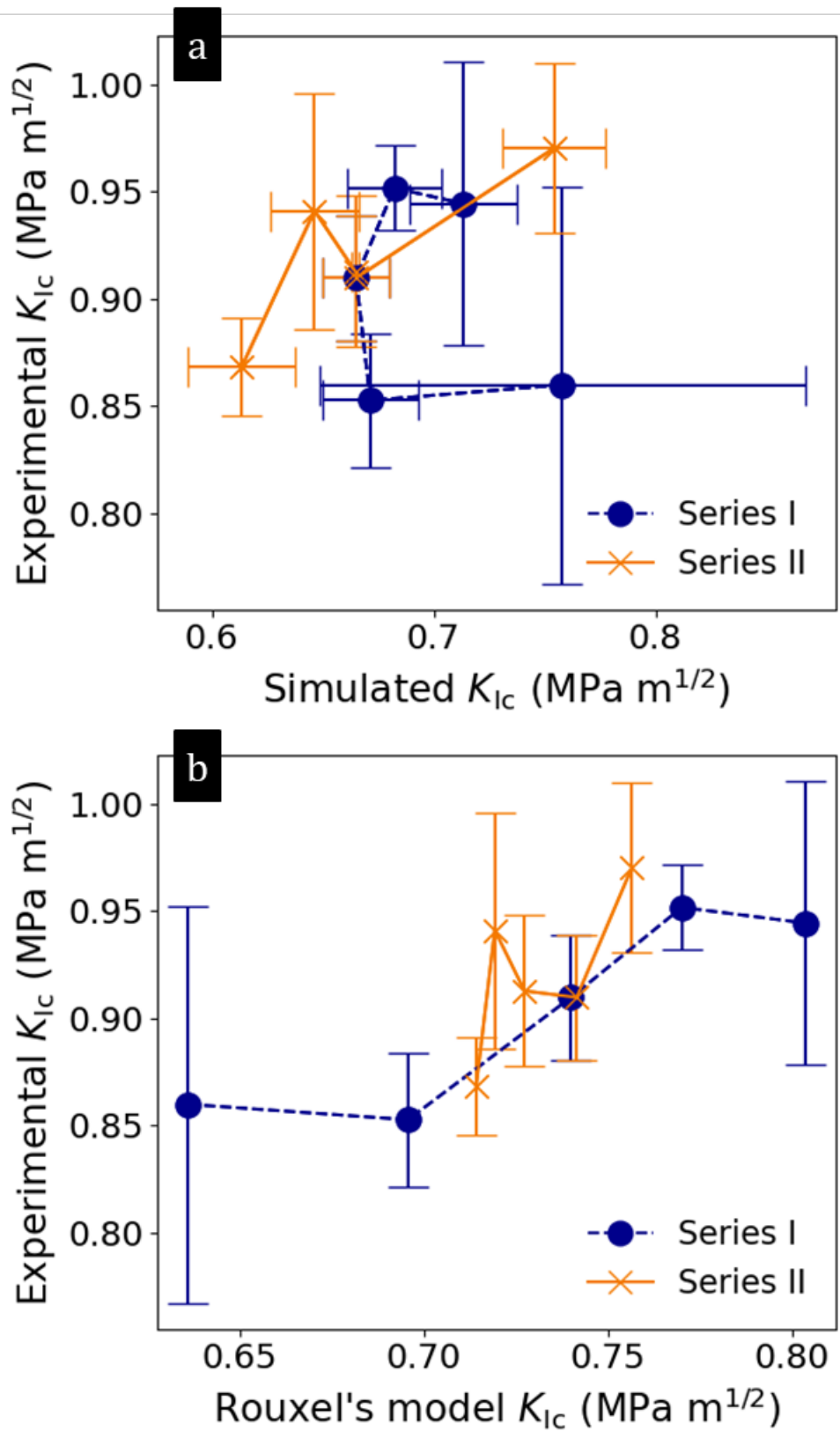

**Figure S7.** (a) Relationship between experimental and MD-simulated fracture toughness ($K_{Ic}$). (b) Relationship between experimental and theoretical fracture toughness as calculated according to Rouxel's model.

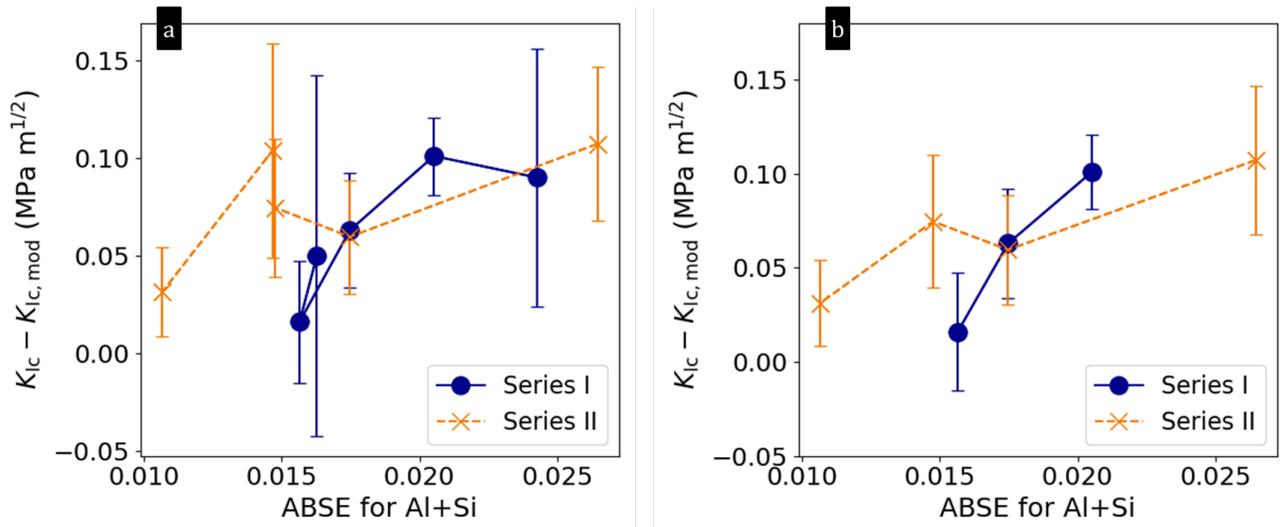

**Figure S8.** (a) Relationship between the difference between the experimental fracture toughness ($K_{Ic}$) and theoretical fracture toughness ($K_{Ic,mod}$). Error bars correspond to the standard deviation of the experimentally measured fracture toughness values. (b) Same relationship as in panel (a) excluding the data points where the standard deviation for the experimental fracture toughness is above $0.05\ MPa \cdot m^{1/2}$.

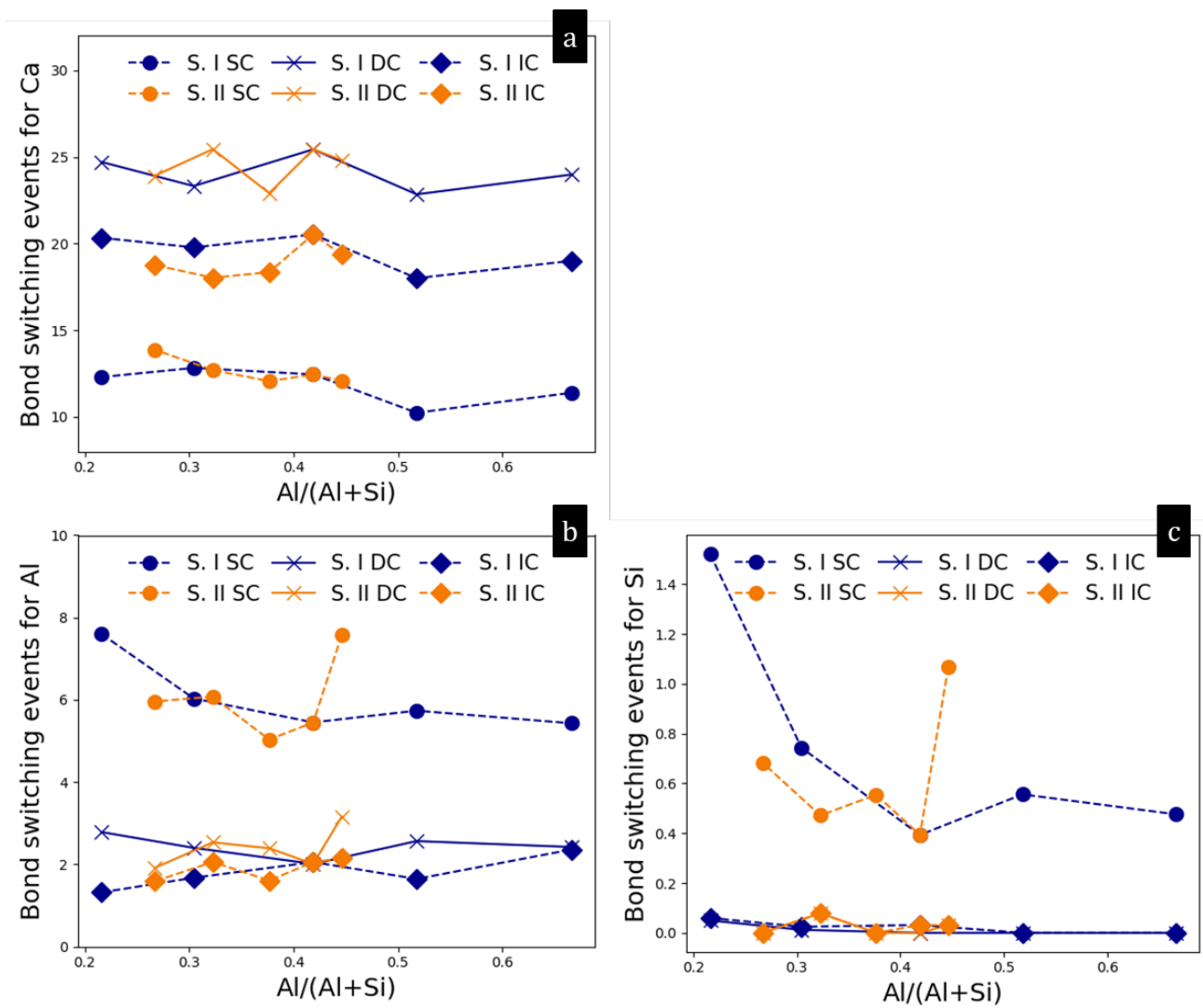

**Figure S9.** Frequency of each bond switching mechanism, i.e., increased coordination number (IC), decreased coordination number (DC) and swapped coordination (SC), for (a) Ca, (b) Al, and (c) Si atoms. Bond switching events are not normalized by the fractions of the elements.